\documentclass[fleqn,12pt,twoside]{article}
\usepackage{espcrc1}
\usepackage{amssymb}  
\usepackage{wasysym}
\usepackage{psfrag}

\usepackage{graphicx}

\newcommand{\AmS}{{\protect\the\textfont2
  A\kern-.1667em\lower.5ex\hbox{M}\kern-.125emS}}

\newcommand{\vp}{{\mathbf p}}    
\newcommand{\vq}{{\mathbf q}}    
\newcommand{\vx}{{\mathbf x}}    
\newcommand{\dm}{{\delta\mu}} 

\title{The crystallography of color superconductivity\thanks{We 
thank
the Kavli Institute for Theoretical Physics in Santa
Barbara for support and hospitality.
Research supported in part by DOE 
cooperative research agreement
\#DF-FC02-94ER40818, by NSF Grant~PHY99-07949, 
and (that of JAB) by a KITP Graduate
Fellowship and DOD NDSEG Fellowship.}}

\author{Jeffrey A. Bowers\footnote{Speaker.}
\address{Center for Theoretical Physics, MIT,
Cambridge, MA 02139, U.S.A.}, 
Krishna Rajagopal\addressmark}
       
\begin{document}

\maketitle



We describe the crystalline phase of color superconducting quark
matter.  This phase may occur in quark matter at 
densities relevant for compact star physics, with possible
implications for glitch phenomena in pulsars~\cite{BowersLOFF}.  We use a
Ginzburg-Landau approach to determine that the crystal has a
face-centered-cubic (FCC) structure~\cite{crystal2}.  Moreover, our
results indicate that the phase
is robust, with gaps, critical temperature, and free energy comparable
to those of the color-flavor-locked (CFL) phase~\cite{crystal2}.  
Our calculations
also predict ``crystalline superfluidity'' in ultracold gases of
fermionic atoms~\cite{AtomicLOFF}.



Cold dense quark matter is a color superconductor \cite{Reviews}.  At
asymptotically high densities, the ground state of QCD with quarks of
three flavors ($u$, $d$, and $s$) is the color-flavor-locked (CFL)
phase \cite{CFL}.  
This phase features a BCS condensate of Cooper pairs of quarks
that includes $ud$, $us$, and $ds$ pairs.  
At intermediate densities, however, the CFL phase can be
disrupted by any flavor asymmetry (such as a chemical potential
difference or a mass difference) that would, in the absence
of pairing, separate the Fermi surfaces.  
In the absence of pairing,
electrically neutral bulk quark matter with
$m_{u,d} = 0$ and $m_s \neq 0$ features a
nonzero electron density ($\mu_e \approx m_s^2/4\mu$) and three disparate
quark Fermi momenta: $p_F^d \approx p_F^u + m_s^2/4\mu$, $p_F^s
\approx p_F^u - m_s^2/4\mu$. (Note that decreasing $\mu$ enhances the
flavor disparity.) 
Accounting for pairing
effects modifies this picture: starting in the CFL phase at large
$\mu$, as we decrease $\mu$ the CFL phase remains 
``rigid''~\cite{neutrality}, with coincident quark Fermi surfaces
and no electrons, until either hadronization or  
a first-order unlocking transition, whichever comes first.
Unlocking occurs at $\mu\approx m_s^2/4\Delta_0$, 
where $\Delta_0$ is the CFL gap. Its value and that of $m_s$ 
are density dependent
and sufficiently uncertain that we do not know whether
unlocking occurs before hadronization. 
Here, we pursue the consequences of assuming that
unlocking occurs first.

\begin{figure}[t]
\begin{minipage}[t]{85mm}
\psfrag{T}{$T$}
\psfrag{mu}[tc][rc]{$\mu_{\mbox{\footnotesize baryon}}$}
\psfrag{hadrons}{hadrons}
\psfrag{crystal}[cc][cc]{XTAL}
\psfrag{cfl}[cc][cc]{CFL}
\psfrag{quarkgluonplasma}{QGP}
\includegraphics[width=74mm]{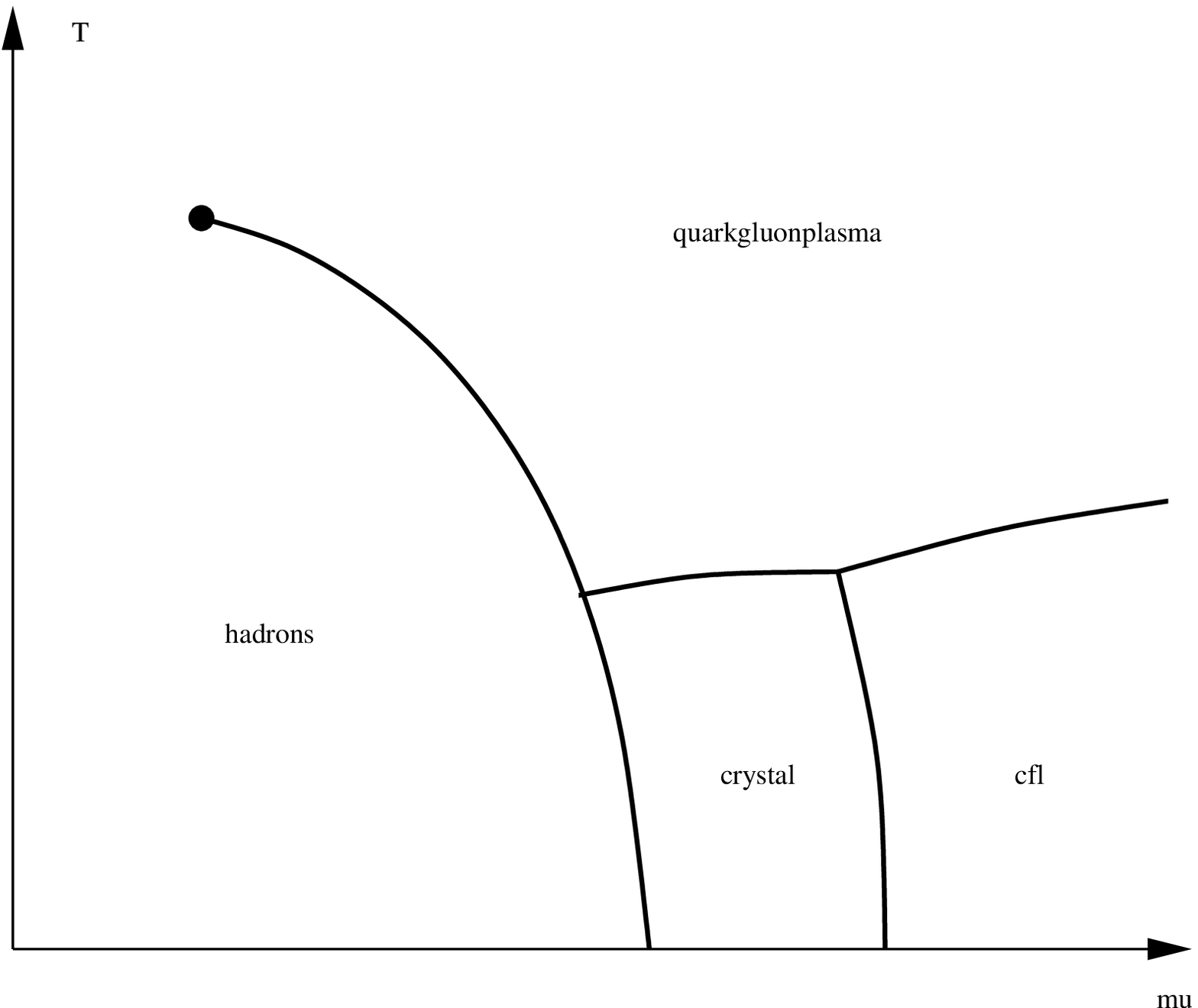}
\vspace{-0.7cm}
\caption{
Schematic QCD phase diagram with quark-gluon
plasma, crystalline (XTAL) and CFL
color superconducting phases, the latter
separated by the
unlocking transition.  
}
\end{minipage}
\hspace{\fill}
\begin{minipage}[t]{70mm}
\psfrag{0}{\small 0}
\psfrag{0.25}{\small 0.25}
\psfrag{0.5}{\small 0.5}
\psfrag{0.75}{\small 0.75}
\psfrag{1}{\small 1}
\includegraphics[width=65mm]{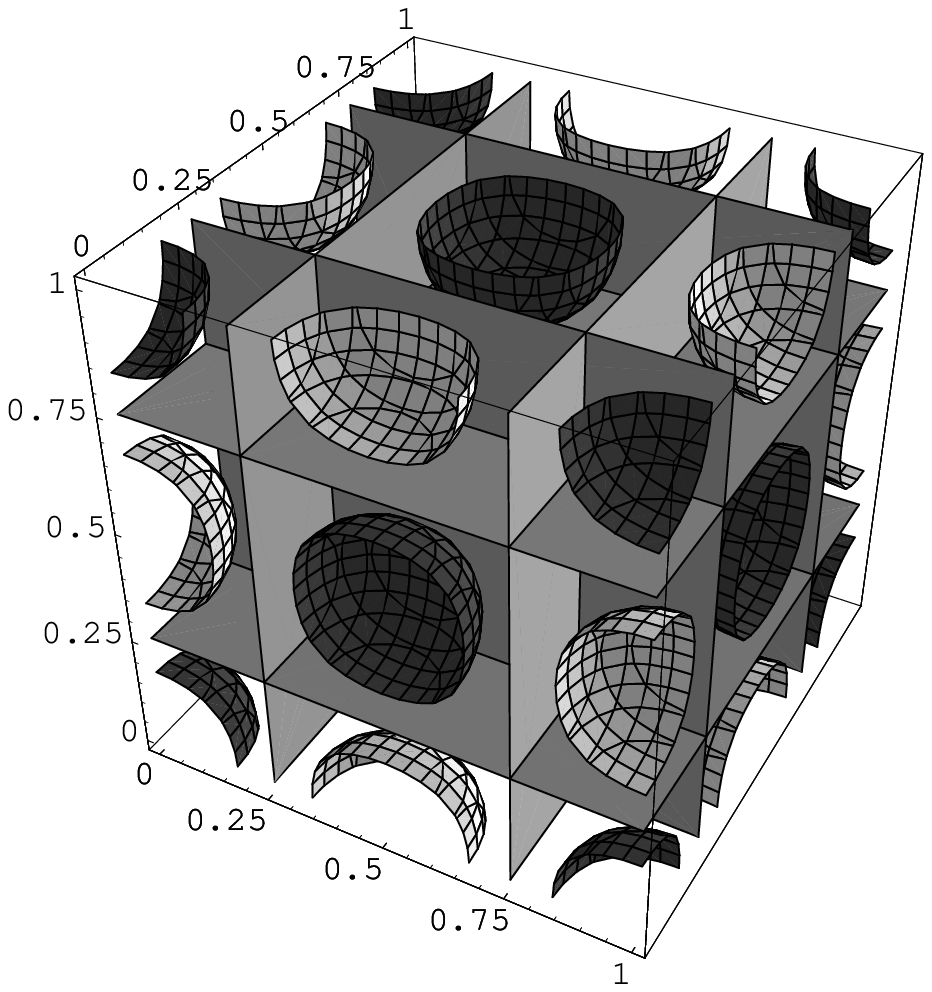}
\vspace{-0.7cm}
\caption{
\label{unitcellfig}
A unit cell of the FCC crystal, with contours where the 
gap function is positive (black), zero (gray),
and negative (white).  
}
\end{minipage}
\vspace{-0.75cm}
\end{figure}

In quark matter below the unlocking transition, pairing can still occur.  One
option is single-flavor pairing ($uu$, $dd$, $ss$), but these
$J=1$ condensates have very small 
gaps~\cite{Schaefer1Flavor}.  Crystalline
color superconductivity is more robust~\cite{BowersLOFF,crystal2}.  
We propose
that unlocked quark matter is in the crystalline phase, which
therefore occupies the window in the QCD phase diagram between the CFL
and hadronic phases (Fig.~1).  
The crystalline phase has the unique
virtue of allowing pairing between quarks with unequal Fermi surfaces.
It was originally described by Larkin, Ovchinnikov, Fulde, and Ferrell
(LOFF) \cite{LOFF} as a novel pairing mechanism for an electron
superconductor 
with a Zeeman splitting between spin-up and
spin-down Fermi surfaces, neglecting all orbital effects
of the magnetic perturbation.
Quark matter is a more natural setting for the LOFF phase, as it
features
a ``flavor Zeeman effect'' with no orbital complications.
Cooper pairs in the LOFF phase
have nonzero total momentum: a quark with momentum $\vp$ is paired
with a quark with momentum $-\vp+2\vq$.
The magnitude $|\vq| \equiv q_0$ is determined by the separation
between Fermi surfaces
while the direction(s) $\hat\vq$ is (are) chosen spontaneously.  
Each
quark in a pair resides near its respective
Fermi surface, so pairs 
can form at low cost
and a condensate 
results.
For a
given $\vq$, the quarks that pair are only those in  
``pairing rings'', one  on each Fermi surface; these circular bands are
antipodal to each other and perpendicular to $\vq$.

If each Cooper pair in the condensate carries the same total momentum
$2\vq$, then in position space the condensate varies like a plane
wave: $\langle \psi(\vx) \psi(\vx) \rangle \sim \Delta
\exp(2i\vq\cdot\vx)$ meaning that translational and rotational symmetry
are broken.  
If the system is unstable to the formation of a
single plane-wave condensate, then we expect that a superposition of
multiple plane waves is more favorable:
$
\langle \psi(\vx) \psi(\vx) \rangle \sim \sum_{\vq, |\vq|=q_0} 
\Delta_\vq e^{2i \vq\cdot\vx}.
$
Each $\Delta_\vq$ correponds to condensation of Cooper pairs with
momentum $2\vq$, i.e.~another pairing ring on each Fermi surface.
As we add more plane waves, we utilize more of the Fermi surface for
pairing, with a corresponding gain in condensation energy.  On the
other hand, the rings can ``interact'' with each other:
condensation in one mode can enhance or deter
condensation in another mode.  We need 
to compare various different choices
for the set of $\hat{\bf q}$'s whose
$\Delta_{\vq}$'s are nonzero; that is, we need to compare
different crystal structures.
To this end,
we calculate the Ginzburg-Landau free
energy $\Omega(\{\Delta_{\vq}\})$ for the LOFF phase.


We consider a simplified model with massless $u$ and $d$ quarks only,
and introduce chemical potentials $\mu_u = \bar\mu - \dm$, $\mu_d =
\bar\mu + \dm$.  In this toy model, we vary $\dm$ by hand to emulate
the effect of splitting the Fermi surfaces as described above.  
(The LOFF phenomenon can equally well be induced by mass
differences between quarks~\cite{massLOFF}.) We
work at zero temperature, as appropriate for compact star physics.
We model the interaction between quarks with a pointlike
four-fermion interaction that mimics single gluon exchange.

At $\dm = 0$, the system forms a BCS superconductor with a gap
$\Delta_0$. 
As we increase $\dm$, the system
exhibits a ``rigidity'' analogous to that of the CFL phase: despite
the imposed stress $\dm$, the gap stays constant and the Fermi
surfaces remain coincident.  When $\dm$ reaches a critical
value $\dm_1 = \Delta_0/\sqrt{2} \approx 0.707 \Delta_0$, the BCS
phase ``breaks'' and the Fermi surfaces separate (this is analogous
to CFL ``unlocking'') .  For $\dm > \dm_1$ a
crystalline phase is possible.  If we consider the simplest LOFF
ansatz, a single plane wave condensate~\cite{BowersLOFF}, 
we find that the LOFF state is
favored in an interval $\dm_1 \leq \dm \leq \dm_2$; at $\dm_1$ there
is a first-order transition from BCS to LOFF, while at $\dm_2 \approx
0.754 \Delta_0$ there is a second-order transition from LOFF to the
normal state (unpaired quarks).  
Each Cooper pair has a total momentum
$2\vq$, where $|\vq| = q_0 \approx 1.20 \dm$; this implies that the 
pairing ring on each Fermi surface has an opening 
angle $\psi_0 \approx 2 \cos^{-1} (\dm/q_0) \approx 67.1^{\circ}$.  

Motivated by the second-order transition at $\dm_2$,
we write a
Ginzburg-Landau free energy valid in the small order parameter 
limit~\cite{crystal2}.
The most general expression consistent with translational and rotational
symmetry is  
\begin{equation}
\Omega
\propto 
\sum_{\vq, |\vq| = q_0} 
\alpha \Delta_\vq^* \Delta_\vq
  + \frac{1}{2} \sum_\square J(\square) \Delta_{\vq_1}^* 
\Delta_{\vq_2} \Delta_{\vq_3}^* \Delta_{\vq_4} 
+ \frac{1}{3} \sum_{\hexagon} K(\hexagon) \Delta_{\vq_1}^* 
\Delta_{\vq_2} \Delta_{\vq_3}^* \Delta_{\vq_4} \Delta_{\vq_5}^* 
\Delta_{\vq_6} + \cdots
\end{equation}
The symbol $\square$ represents a set of four equal-length vectors
$(\vq_1, \vq_2, \vq_3, \vq_4)$, $|\vq_i| = q_0$, with $\vq_1 - \vq_2
+\vq_3 - \vq_4 = 0$, forming a closed figure.
Similarly, the
symbol $\hexagon$ represents a set of six equal-length vectors
$(\vq_1, \ldots, \vq_6)$ with $\vq_1 - \vq_2 + \vq_3 - \vq_4 - \vq_5 +
\vq_6 = 0$.
The quadratic coefficient $\alpha$ changes sign at $\dm_2$ showing the
onset of the LOFF plane-wave instability: $\alpha \approx (\dm -
\dm_2)/\dm_2$.  Analysis of the quadratic term shows that 
all modes on the sphere $|\vq| = q_0$ 
become unstable at $\dm<\dm_2$, and that at larger $\dm$ it
is these modes that are the least stable.
The quartic and sextic coefficients $J$ and $K$ characterize the
interactions between modes, and thus determine the crystal structure.  
For candidate crystal structures with 
all $\Delta_\vq$'s equal in magnitude, 
we can evaluate aggregrate Ginzburg-Landau
quartic and sextic coefficients $\beta$ and $\gamma$ as sums over all rhombic
and hexagonal combinations of the $\vq$'s:  
$\beta = \sum_{\square} J(\square)$, $\gamma = \sum_{\hexagon} K(\hexagon)$.  
Then for a crystal with $P$ plane
waves we obtain 
$
\Omega(\Delta) \propto P \alpha \Delta^2 + 
\frac{1}{2} \beta \Delta^4 + \frac{1}{3} \gamma \Delta^6 + \cdots
$
and we compare crystals by calculating $\beta$
and $\gamma$ to find the structure with the lowest $\Omega$.


We have analyzed many candidate crystal structures
in \cite{crystal2};  here, we summarize the
lessons learned.  Recall that each $\vq$
corresponds to a pairing ring on each Fermi surface with opening angle
$\psi_0 \approx 67.1^{\circ}$.  The first lesson
is that adding more pairing rings
lowers the free energy only as long as rings do not intersect.
No more than nine rings can be
arranged on the Fermi surface without any overlaps.  The
second lesson is that ``regular''
structures (those with nonintersecting rings whose
wave vectors 
can be assembled into closed figures in many ways)
are combinatorically favored.
There are no particularly regular nine-wave structures, but there is a 
very regular eight-wave
structure: eight wave vectors $\vq$ pointing towards the eight corners
of a cube, forming the eight shortest vectors in the reciprocal
lattice of a face-centered-cubic (FCC) crystal.  This FCC crystal has
by far the lowest Ginzburg-Landau free energy of all the structures
we considered.  The gap function is
$
\langle \psi(\vx) \psi(\vx) \rangle 
\sim 
\Delta 
[
\cos\frac{2\pi}{a} (x+y+z) + \cos\frac{2\pi}{a} (x-y+z) 
+ \cos\frac{2\pi}{a}(x+y-z)
+ \cos\frac{2\pi}{a} (-x+y-z) 
]
$
with lattice constant $a = \sqrt{3} \pi/|\vq| \approx 6.012/\Delta_0$.  
A unit cell is shown in Fig.~2.  
The effective field theory for the phonons
of this  crystal has recently been constructed~\cite{LOFFphonon}.  

The Ginzburg-Landau free energy for the FCC structure has 
coefficients $\beta$ and $\gamma$ that are large and negative.
This guarantees that the phase transition to the crystalline phase
cannot be a second-order transition at $\dm=\dm_2$ and must instead
be a strongly first-order transition at some $\dm=\dm_*\gg \dm_2$,
at which the magnitude of the
gap jumps from zero to a large nonzero value.  
Were we to push beyond sextic order
to obtain a bounded Ginzburg-Landau free energy, the resulting
calculation of the gap and ground state free energy 
would be quantitatively uncontrolled because 
the order parameter is nowhere nonzero and small.
Qualitatively, the large negative coefficients indicate a robust
crystalline phase, with gaps and free energy comparable to those of
the BCS phase. Furthermore, because the transition
to the crystalline phase occurs at $\dm_*\gg \dm_2$,
the crystalline phase is favored over a wide $\dm$ interval.
Our results
thus indicate that wherever unlocked quark matter occurs, 
it will be in the crystalline phase
with an FCC crystal structure.

The power of the Ginzburg-Landau formulation is that it organizes
the calculation in such a way that 
simple qualitative lessons about what features make
a crystal structure favorable emerge. 
We immediately learn that, among all the possibilities,
the FCC structure of Fig.~2
is the best choice.  This is the
first time (in either the condensed matter or the QCD
incarnations of this subject) that sufficiently general 
three dimensional crystal structures have been compared.
The structure that we find to be most favorable, although
simple with hindsight, was never considered
in the classic condensed matter physics papers.
Unfortunately, but not unusually, the Ginzburg-Landau calculation
predicts a first-order phase transition and in so doing predicts
its own quantitative demise.  With a qualitative understanding
of what crystal structure wins (that is, with confidence 
in what ansatz to make)
we are currently doing the variational calculation needed
to determine 
the gap, free energy, and $\dm_*$ quantitatively
within our toy model.  
Going beyond our toy model, a full three-flavor analysis is
desirable, with $ud$, $us$ and $ds$ crystalline condensates and a
strange quark mass.  
In an atomic physics
context, the FCC condensate may be detectable as a ``crystalline
superfluid'' in an ultracold gas of fermionic atoms \cite{AtomicLOFF}.
In an astrophysics context, with the crystal structure known a calculation
of the pinning force can now proceed, in order to determine
whether (some) observed pulsar glitches 
originate from pinning of rotational vortices at intersections of
crystal nodal planes.  

\end{document}